\documentstyle[eqsecnum,preprint,aps]{revtex}

\begin{document}
\draft
\begin{titlepage}
\baselineskip .25in
\begin{flushright}
% preprint
WU-AP/49/95

gr-qc/9506068
% $~$
\end{flushright}
\begin{center}
\vskip 1cm
{\large{\bf Dynamics of Topological Defects and Inflation}}

\vskip 0.5cm
{\sc Nobuyuki Sakai\footnote{electronic mail: sakai@cfi.waseda.ac.jp},
Hisa-aki Shinkai\footnote{electronic mail: shinkai@cfi.waseda.ac.jp},
Takashi Tachizawa}\footnote{electronic mail: 63L507@cfi.waseda.ac.jp}
\\and {\sc Kei-ichi Maeda}\footnote{electronic mail:
maeda@cfi.waseda.ac.jp}

\vskip 0.3cm
\baselineskip = 24pt
{\it Department of Physics, Waseda University, 3-4-1 Okubo,
Shinjuku-ku, Tokyo 169} \end{center}

\vskip 1cm
\begin{center}
{\bf Abstract}
\end{center}

% preprint
\baselineskip = 18pt
We study the dynamics of topological defects in the context of
``topological inflation" proposed by Vilenkin and Linde independently.
Analysing the time evolution of planar domain walls and of global
monopoles, we find that the defects undergo  inflationary expansion if
$\eta\stackrel{>}{\sim}0.33m_{Pl}$, where $\eta$ is the vacuum
expectation value of the Higgs field and $m_{Pl}$ is the Planck mass.
This result confirms the estimates by Vilenkin and Linde. The critical
value of $\eta$ is independent of the coupling constant $\lambda$ and
the initial size of the defect. Even for defects with an initial size
much greater than the horizon scale, inflation does not occur at all
if $\eta$ is smaller than the critical value. We also examine the
effect of gauge fields for static monopole solutions and find that the
spacetime with a gauge monopole has an attractive nature, contrary to
the spacetime with a global monopole. It suggests that gauge fields
affect the onset of inflation.

\vskip 1cm
\noindent
% preprint
\begin{center}
To appear in {\it Physical Review D}
\end{center}

% PACS number(s): 04.20.-q, 98.80.Cq
\vfill

\end{titlepage}

% preprint
\baselineskip = 18pt
\section{Introduction}

Vilenkin \cite{topo-vilenkin} and Linde \cite{topo-linde}
independently pointed out that topological defects can be seeds for
inflation. The basic idea of this  ``topological inflation" is as
follows. Suppose that we have the Higgs field $\Phi^a ~(a=1, \cdots,
N)$, whose potential is
\begin{equation}\label{pote}
V(\Phi)= {1\over 4}\lambda(\Phi^2-\eta^2)^2, ~~
\Phi\equiv\sqrt{\Phi^a\Phi^a},
\end{equation}
where $\eta$ is the vacuum expectation value and $\lambda$ is a
coupling constant. This model gives rise to different types of
topological defects, {\it i.e.}, domain walls for $N=1$, cosmic
strings for $N=2$ and monopoles for $N=3$. Because the center of a
defect is in the false vacuum state ($\Phi=0$), we can expect
that inflation occurs near $\Phi=0$ under some condition. Vilenkin and
Linde claimed that topological defects expand exponentially if and
only if
\begin{equation}\label{etapl} \eta>O(m_{Pl})
\end{equation}
for the following reasons.

\begin{enumerate}
\item[(i)] If the size of the false vacuum region is greater than the
horizon size, it is natural to assume that this region undergoes
inflationary expansion. The thickness of a domain wall in a flat
spacetime is given by
\begin{equation}
\delta_0={\sqrt{2}\over\sqrt{\lambda}\eta},
\end{equation}
and the horizon size corresponding to the vacuum energy $V(0)$ is
\begin{equation}
H_0^{-1}\equiv\biggl[{8\pi\over3m_{Pl}^{~2}}V(0)\biggr]^{-\frac12}
=\sqrt{{3\over2\pi\lambda}}{m_{Pl}\over\eta^2}.
\end{equation}
Then the condition $\delta_0>H_0^{-1}$ implies $\eta\stackrel{>}{\sim}
m_{Pl}$ \cite{topo-vilenkin}.

\item[(ii)] New inflation or chaotic inflation occurs in
Friedmann-Robertson-Walker spacetime if the slow-rolling condition
($|\ddot{\Phi}|\ll3H|\dot\Phi|$ and $\dot\Phi^2/2\ll V(\Phi)$) is
satisfied. In the model (\ref{pote}) this condition leads to $\eta\gg
m_{Pl}/\sqrt{6\pi}$ \cite{topo-vilenkin,topo-linde}.

\item[(iii)] Gravitational effect becomes important if the Schwarzschild
radius $R_g=2M/m_{Pl}^2$ is comparable to $\delta_0$, where
$M\approx(4\pi/3)\delta_0^3V(0)$. This condition is reduced to
$\eta\stackrel{>}{\sim}\sqrt{3/8\pi}m_{Pl}$
\cite{topo-vilenkin,topo-linde}.

\item[(iv)] In the case of a global string or a global monopole there
is a
deficit angle or a solid deficit angle in the static solution. For
each defect, this exceeds $2\pi$ or $4\pi$ if $\eta>m_{Pl}$, which
indicates no static solution with an asymptotically flat region
\cite{topo-vilenkin}.

\item[(v)] It was shown in \cite{breit1,lee,breit2} that nonsingular
static gauge monopole solutions exist only if $\eta\stackrel{<}{\sim}
m_{Pl}$; otherwise, the  monopole becomes dynamic and may become a
Reissner-Nordstr$\ddot{{\rm o}}$m black hole. Linde speculated that in
this case the central region of the monopole expands exponentially
because the condition, $\eta\stackrel{>}{\sim} m_{Pl}$, is simultaneously
a condition of inflation (the slow-rolling condition in (ii)).

\end{enumerate}

In this paper we study topological inflation in detail by numerical
analysis. On this subject, we need to clarify the following points.

\begin{enumerate}
\item[1.] Although the condition for inflation to occur, (\ref{etapl}),
looks plausible, it is based on discussions which did not show how the
defects really inflate. It is important to verify these arguments by
numerical analysis which comprehensively includes the gravitational
effect.

\item[2.] Vilenkin and Linde assumed the initial size of defects to be
$\delta_0$, which stems from the static solution without gravitational
effect. We agree that $\delta_0$ implies the typical size of the
defects even in curved spacetime. In the beginning of the phase
transition, however, the scalar field takes its phase randomly and
global distribution may be chaotic. Therefore, it is worth studying
the dynamics of the defects with various initial sizes. Particularly,
our interest is the fate of the topological defect when the initial
size is greater than the horizon scale but $\eta$ is not so large. One
may conceive that such defects also inflate.

\item[3.] The effect of gauge fields is unclear. Linde claimed that
gauge  fields do not affect inflation of strings or monopoles, because
they exponentially decrease during inflation. We agree that, once
inflation occurs, their effect gets smaller and smaller. However, it
remains an unsettled question whether or not gauge fields affect the
onset of inflation. \end{enumerate}

In order to answer the first and the second questions, we investigate
the evolution of planar domain walls and global monopoles in \S 2. As
to the third question, we discuss static monopole solutions in \S 3.
\S 4 is devoted to conclusions. In this paper we use the units
$c=\hbar=1$.

\section{Evolution of Planar Domain Walls and Global Monopoles}

In what follows we numerically analyse the time-evolution of domain
walls ($N=1$) in a plane  symmetric spacetime and of global monopoles
($N=3$) in a spherically symmetric spacetime. The Einstein-Higgs
system is described by the action,
\begin{equation}\label{EHaction}
  S=\int d^4 x \sqrt{-g} \left[\frac{m_{Pl}^{~2}}{16\pi}{\cal R}
     -\frac12(\partial_{\mu}\Phi^a)^2-V(\Phi)\right].
\end{equation}
The field equations derived from (\ref{EHaction}) are presented in
Appendix.

For the case of domain walls, we assume that the metric has a form,
\begin{equation}\label{plane}
ds^2=-dt^2+A^2(t,|x|)dx^2 + B^2(t,|x|)(dy^2 + dz^2).
\end{equation}
As an initial configuration of the scalar field, it seems natural to
use the same functional form as the static solution in a flat
spacetime, $\Phi_{flat}(x)\equiv\eta\tanh (x/\delta)$, where $\delta$
is a parameter of the initial width of the domain wall. In a curved
spacetime, however, if we consider a single domain wall in a vacuum
background spacetime, the spacetime becomes closed \cite{ipser}. There
is no far region from the wall. For this reason we adopt the periodic
boundary condition and accordingly change the functional form for each
section slightly into
\begin{eqnarray}
\Phi(t=0,x) = \left\{
\begin{array}{ll}
\eta\biggl[{x\over\delta}-{5\over4}\Bigl({8\over15}{x\over\delta}
\Bigr)^3+
{3\over8}\Bigl({8\over15}{x\over\delta}\Bigr)^5 \biggr]& ~~~
(0\le{x\over\delta}\le{15\over8}) \\ \eta & ~~~
({15\over8}\le{x\over\delta}\le2). \end{array} \right.
\end{eqnarray}
The above polynomial has been determined from the following
assumptions: (1) 5th-order odd polynomial function of $x$; (2) The
first term agrees with that of $\eta\tanh (x/\delta)$; (3) It connects
the constant function $\Phi(x)=\eta$ at the point where
$\Phi'\equiv\partial\Phi/\partial x =\Phi''=0$. For later convenience,
we define a parameter of the initial width of a wall as
$c\equiv\delta/\delta_0$.

One may wonder if our results really show generic properties of domain
walls, because the global structure of a spacetime depends on how we
choose boundary conditions. We will discuss this point in the final
section.

For the case of global monopoles, we assume a spherically symmetric
spacetime:
\begin{equation}\label{sphere}
ds^2=-dt^2+A^2(t,r)dr^2+B^2(t,r)r^2(d\theta^2+\sin^2\theta d\varphi^2).
\end{equation}
For the scalar field, we adopt the hedgehog ansatz:
\begin{equation}\label{hedgehog}
\Phi^a=\Phi(t,r)\hat r^a\equiv
\Phi(t,r)(\sin\theta\cos\varphi,\sin\theta\sin\varphi,\cos\theta).
\end{equation}
As an initial configuration of the scalar field, we suppose
\begin{equation}
\Phi(t=0,r)=\Phi_{flat}\Bigl({r\over c}\Bigr),
\end{equation}
where $\Phi_{flat}(r)$ is a static solution in a flat spacetime, which
is obtained only numerically, and $c$ is an initial size parameter of
a monopole normalized by that of the static solution.

As to the initial value of $\dot\Phi\equiv \partial\Phi/\partial t$,
we suppose $\dot\Phi=0$ in both the cases. Following the method shown
in the Appendix, we solve the constraint equations for setting initial
data and the dynamical equations for time evolution.

Our numerical results are summarized in Figs.1-5. In Fig.1 we show
examples of the time-evolution of the scalar field, which correspond
to a stable domain wall ((a) $\eta=0.2m_{Pl}$) and an inflating wall
((b) $\eta=0.6m_{Pl}$), respectively. The abscissa, $X$, is defined as
a proper distance along the $x$-axis from the origin. The corresponding
results for monopoles are presented in Fig.2. In the case of (b) we
see that the bottle-neck structure appears, but we do not find an
apparent horizon of a black-hole; this is not a worm-hole. (As for an
apparent horizon, see the last part of Appendix.) In Fig.3 we draw
trajectories of the boundaries of a domain wall and a monopole. Here
we define the boundary of the defect as $X_b(t)=X$ or $R_b(t)=Br$ at
the position of $\Phi=\eta/2$. Fig.3 indicates that a domain wall or a
monopole expands if $\eta\ge0.4m_{Pl}$, while it remains stable if
$\eta\le0.3m_{Pl}$. We survey for $0.3m_{Pl} \le\eta\le 0.4m_{Pl}$ and
$10^{-4}\le\lambda\le10$ closely, and find the critical value of $
\eta$ is around $0.33m_{Pl}$, regardless of $\lambda$. This result
supports the estimates of Vilenkin and Linde.

By varying $c$, we also investigate how the initial size of a domain
wall or a  monopole influences its dynamics. Fig.4 indicates that the
final behavior of a domain wall or a monopole is determined only by
$\eta$. In the case of (a) or (b) ($\eta=0.2m_{pl}$), even if the
initial size is much larger than $H_0^{-1}$, the configuration
eventually approaches to $\Phi_{flat}(X)$ or $\Phi_{flat}(Br)$. We
analyse whether or not the cosmological horizon appears or not by
searching for an apparent horizon, because $H_0^{-1}$ is not exactly
the cosmological horizon. Fig.5 shows that an apparent horizon really
appears and the surface of the shrinking monopole ($\eta=0.2m_{Pl},
c=10$) crosses it. Thus we see that, even for defects with an initial
size much greater than the horizon scale, inflation does not occur if
$\eta$ is smaller than the critical value.

\section{Effect of Gauge Fields for Static Monopole Solutions}

In this section we describe the effect of gauge fields for static
monopole solutions. We consider the SU(2) Einstein-Yang-Mills-Higgs
system, which is described by
\begin{equation}
  S=\int d^4 x \sqrt{-g} \left[\frac{m_{Pl}^{~2}}{16\pi}{\cal R}
     -\frac14(F^a_{\mu\nu})^2
     -\frac12(D_{\mu}\Phi^a)^2-V(\Phi)\right],
\end{equation}
with
\begin{equation}
F^a_{\mu\nu} \equiv \partial_{\mu}A^a_{\nu}-\partial_{\nu}A^a_{\mu}
  -e\epsilon^{abc}A^b_{\mu}A^c_{\nu},~~
D_{\mu}\Phi^a\equiv\partial_{\mu}\Phi^a+e\epsilon^{abc}
A^b_{\mu}\Phi^c,
\end{equation}
where $A^a_{\mu}$ and $F^a_{\mu\nu}$ are the SU(2) Yang-Mills field
potential and its field strength, respectively. $D_{\mu}$ is the
covariant derivative. A static and spherically symmetric spacetime is
described  as
\begin{equation}\label{static}
ds^{2}= -\Bigl(1-\frac{2m_{Pl}^{-2}M(R)}{R}\Bigr)e^{-2\alpha(R)}dT^{2}
+\Bigl(1-\frac{2m_{Pl}^{-2}M(R)}{R}\Bigr)^{-1}dR^{2}+R^{2}(d\theta^{2}
+\sin^{2}\theta d\varphi^{2}).
\end{equation}
Under the 't Hooft-Polyakov ansatz, the Yang-Mills potential is
written as
\begin{equation}
A^a_i=\omega^c_i\epsilon^{cab}\hat r^b{1-w(R)\over eR},~~ A^a_0=0,
\end{equation}
where $\omega^a_i$ is a triad.

Breitenlohner et al. systematically surveyed regular monopole solutions
in this system \cite{breit1}. According to Fig.6 in \cite{breit1},
static solutions cease to exist when $\eta$ becomes larger than a
critical value $\eta_{cr}$, which has a little dependence on $\lambda/e^2$
: $0.20m_{Pl}<\eta_{cr}(\lambda/e^2) <0.39m_{Pl}$ for
$\infty\ge\lambda/e^2\ge0$. (The relations between the parameters in
this paper and those in \cite{breit1} are given by
$\eta/m_{Pl}=\sqrt{4\pi}\alpha$ and $\lambda/e^2=\beta^2/2$.) The exact
value of $\eta_{cr}$ for $\lambda/e^2\rightarrow\infty$ was analytically
obtained \cite{breit2}:
$\eta_{cr}(\lambda/e^2\rightarrow\infty)=m_{Pl}/\sqrt{8\pi}$. We also
examine this monopole system and confirm their results.

{}From the two facts of the disappearance of static solutions for large
$\eta$ and of the inflation of a global monopole for large $\eta$, we
might expect that gauge monopoles also inflate for large $\eta$. In
order to see the effect of gauge fields, we calculate the acceleration
of test particles, which is governed by the geodesic equation in the
coordinate system (\ref{static}):
\begin{equation}\label{accel}
{d^2R\over d\tau^2}=-{e^{2\alpha(R)}\over2}{d\over dR}
\biggl[e^{-2\alpha(R)}
\Bigl\{1-{2m_{Pl}^{-2}M(R)\over R}\Bigr\}\biggr],
\end{equation}
where $\tau$ is the proper time and we have assumed
$dR/d\tau=d\theta/d\tau=d\varphi/d\tau=0$. Although we cannot discuss
the dynamics of a static spacetime by its metric itself, the
acceleration of test particles gives us some information about the
spacetime structure. For example, in de Sitter spacetime
($M(R)=m_{Pl}^2\bar H^2R^3/2, ~ \bar H=$ const., $\alpha(R)=0$) and in
Schwarzschild spacetime ($M(R)=\bar M=$ const., $\alpha(R)=0$),
(\ref{accel}) is reduced to $d^2R/d\tau^2=\bar H^2R>0$ and
$d^2R/d\tau^2=-m_{Pl}^{-2}\bar M/R^2<0$, respectively. The sign of
acceleration indicates whether the spacetime is repulsive $(+)$ or
attractive $(-)$. In Fig.6 we present examples of the acceleration in
spacetimes with a global monopole and with a gauge monopole. The
figures show that the two spacetimes have different properties: the
spacetime with a global monopole has a repulsive nature \cite{harari},
while the spacetime with a gauge monopole has an attractive nature.
Neither sign of acceleration changes even for large $\eta$. Although
we have analysed only static spacetimes, our results indicates that
there may be another possibility that the monopole collapses and
becomes a black hole.

\section{Discussions}

We have studied the dynamics of topological defects numerically. We
have examined three questions on ``topological inflation" and found
the following results.

The first subject was simply verifying the arguments of Vilenkin and
Linde. Our results support their discussions completely and we obtain
more exactly the critical value of $\eta$ which determines whether
defects inflate or not. We found that, if
$\eta\stackrel{>}{\sim}0.33m_{Pl}$ is satisfied, planar domain walls
and global monopoles inflate.

The second was to examine the dynamics of defects of various initial
size. Our results show that in the case of planar domain walls and
global monopoles, only $\eta$ determines whether or not it expands.
Even if the initial size of the defect is much greater than the
horizon scale, it shrinks and approaches a stable configuration if
$\eta$ is less than the critical value. Some readers may feel this
result is surprising, but we can reasonably interpret it as follows.
First, we found in our analysis that the boundary of a defect
($X_b(t)$ or $R_b(t)$) can be a spacelike hypersurface, which was also
pointed out by Vilenkin \cite{topo-vilenkin}. Thus there is no reason
that the ``horizon scale" prevents a defect from shrinking. Secondly,
if $\eta\ll m_{Pl}$, the slow-rolling condition is not satisfied. This
condition is expressed only by $\eta$, regardless of $\lambda$ and the
horizon scale. We therefore conclude that, although the statement
{\it``If the size of a defect is greater than the horizon scale, it
inflates.}" sounds true, in a strict sense it is not.

Let us summarize the main points: first, the condition of inflation
for domain and for global monopoles are the same; secondly, the
condition does not depend on the initial size. From these two facts,
we may understand that whether inflation occurs or not is determined
not by the global structure of a spacetime but by the local ``slow-rolling"
condition; our results indicate that slow-rolling condition is a
necessary and sufficient condition for topological inflation. Therefore,
although we have assumed a plane symmetric spacetime and the periodic
boundary condition for a domain wall system, it seems reasonable to
conclude that our results show generic properties of domain walls.
Similarly, we may extend our results to the case of global strings.

The final subject we have investigated is the effect of gauge fields.
Comparing the gauge monopole static solution with the global monopole
one, we find that these spacetimes have different characters: one is
attractive and the other is repulsive. Therefore, gauge fields work to
obstruct inflation and their effect cannot be ignored when we discuss
the onset of inflation. It may be interesting to study time-dependent
gauge monopoles, which would give us a definite answer to this issue.

\acknowledgements

N.S. would like to thank K. Nakao for valuable advice about numerical
procedure. Thanks are also due to W. Rozycki for correcting the
manuscript. T.T. is thankful to JSPS for financial support.
This work was supported partially by the Grant-in-Aid for Scientific
Research Fund of the Ministry of Education, Science and Culture
(No.07740226, No.07854014, No.06302021 and No.06640412), by the
Grant-in-Aid for JSPS Fellow (No. 053769), and by a Waseda University
Grant for Special Research Projects.

\appendix
\section{Field Equations and Numerical Method}

In this Appendix we explain how we solve the field equations for
planar domain walls and for global monopoles. At the end we also
summarize how we search for an apparent horizon in a global monopole
system.

The variation of (\ref{EHaction}) with respect to $g_{\mu\nu}$ and
$\Phi^a$ yield the Einstein equations,
$$
G_{\mu\nu}\equiv{\cal R}_{\mu\nu}-\frac12g_{\mu\nu}{\cal R}={8\pi\over
m_{Pl}^{~2}}T_{\mu\nu},
$$
\begin{equation}\label{ein}
T_{\mu\nu}=\partial_{\mu}\Phi^a\partial_{\nu}\Phi^a
-g_{\mu\nu}\Bigl[\frac12(\partial_{\sigma}\Phi^a)^2+V(\Phi)\Bigr],
\end{equation}
and the scalar field equations,
\begin{equation}\label{scalar}
\kern1pt\vbox{\hrule height 1.2pt\hbox{\vrule width1.2pt\hskip 3pt
\vbox{\vskip6pt}\hskip 3pt\vrule width 0.6pt}\hrule height
0.6pt}\kern1pt \Phi^a=\frac{\partial V(\Phi)}{\partial\Phi^a}.
\end{equation}

\vskip.5cm \noindent
(1) Planar domain wall system

Using the metric (\ref{plane}), we write down the field equations
(\ref{ein}) and (\ref{scalar}) as
\begin{equation}\label{hcdw}
-G^0_0 \equiv K^2_2(3K^2_2-2K)-{2B''\over A^2B}-{B'^2\over
A^2B^2} +{2A'B'\over A^3B} = {8\pi\over m_{Pl}^{~2}}
\Bigl({\dot\Phi^2\over2}+{\Phi'^2\over2A^2}+V\Bigr),
\end{equation}
\begin{equation}\label{mcdw}
\frac12G_{01} \equiv {K^2_2}'+{B'\over
B}(3K^2_2-K)={4\pi\over m_{Pl}^{~2}}\dot\Phi\Phi',
\end{equation}
\begin{equation}\label{dotkdw}
\frac12(G^1_1+G^2_2+G^3_3-G^0_0)
\equiv \dot K-{K^1_1}^2-2{K^2_2}^2 =
{8\pi\over m_{Pl}^{~2}}(\dot\Phi^2-V)
\end{equation}
\begin{equation}\label{dotk22dw}
-{\cal R}^2_2-G^0_0 \equiv \dot K^2_2 + {B'^2\over2A^2B^2} -
{3\over2}K^2_2
= {4\pi\over m_{Pl}^{~2}}\Bigl({\dot\Phi^2\over2}+{\Phi'^2\over2A^2}
-V\Bigr)
\end{equation}
\begin{equation}\label{ddotphidw}
\ddot\Phi-K\dot\Phi-{\Phi''\over A^2}-\Bigl(-{A'\over A}+{2B'\over B}
\Bigr){\Phi'\over A^2}
+{dV\over d\Phi}=0
\end{equation}
where we have introduced the extrinsic curvature tensor of $t=$
constant hypersurface, $K_{ij}$, whose components are given by
\begin{equation}\label{kdef}
K^1_1=-{\dot A\over A},~~~ K^2_2~(=K^3_3)~=-{\dot B\over B},
\end{equation}
and denoted its trace by $K\equiv K^i_i$.

In order to set up initial data, we have to solve the Hamiltonian
constraint equation (\ref{hcdw}) and the momentum constraint equation
(\ref{mcdw}). We assume the homogeneous and isotropic curvature,
\begin{equation}\label{kconst}
{K\over3}=K^1_1=K^2_2={\rm const.},
\end{equation}
which makes (\ref{mcdw}) trivial, and the conformally flat spatial
gauge, $A=B$. Because we have adopted the periodic boundary condition
such that the period is $4\delta$, and assumed reflection symmetry, we
must keep the condition $B'(t,x=0)=B'(t,x=2\delta)=0$. By changing $K$
as a shooting parameter, we iteratively integrate (\ref{hcdw}) until
the above condition is satisfied.

Now we have five dynamical variables: $A,~B,~K,~K^2_2$ and $\Phi$.
(\ref{kdef}), (\ref{dotkdw}) and (\ref{ddotphidw}) provide the next
time-step of $A,~B,~K$ and $\Phi$. For the time-evolution of $K^2_2$,
we use (\ref{dotk22dw}) only at $x=0$, and then integrate (\ref{mcdw})
in the $x$-direction to obtain other values of $K^2_2$. In this way we
have reduced spatial derivatives appearing in the equations, which may
become seeds for numerical instability.

\vskip.5cm \noindent
(2) Global monopole system

In a method similar to that applied for the domain wall system, we
solve the field equations for the monopole system. Here we present the
equations and only give comments on some differences. Under the
assumption of the metric (\ref{sphere}) and the hedgehog ansatz
(\ref{hedgehog}), we write down the field equations (\ref{ein}) and
(\ref{scalar}) as \begin{eqnarray}\label{hcmp}
-G^0_0&\equiv&K^2_2(3K^2_2-2K)-{2B''\over A^2B}-{B'^2\over A^2B^2}
+{2A'B'\over A^3B}-{6B'\over A^2Br}+{2A'\over A^3r} -{1\over
A^2r^2}+{1\over B^2r^2} \nonumber\\
&=& {8\pi\over m_{Pl}^{~2}}
\Bigl({\dot\Phi^2\over2}+{\Phi'^2\over2A^2}+{\Phi^2\over
B^2r^2}+V\Bigr), \\ \frac12G_{01}&\equiv& {K^2_2}'+\Bigl({B'\over
B}+{1\over r}\Bigr)(3K^2_2-K) ={4\pi\over m_{Pl}^{~2}}\dot\Phi\Phi'.
\label{mcmp}\end{eqnarray} \begin{equation}
\frac12(G^1_1+G^2_2+G^3_3-G^0_0) \equiv \dot K-{K^1_1}^2-2{K^2_2}^2 =
{8\pi\over m_{Pl}^{~2}} (\dot\Phi^2-V)
\end{equation}
\begin{equation}
\ddot\Phi-K\dot\Phi-{\Phi''\over A^2}-\Bigl(-{A'\over A}+{2B'\over B}
+{2\over r}\Bigr){\Phi'\over
A^2} +{2\Phi\over B^2r^2}+{dV\over d\Phi}=0
\end{equation}

In this system the homogeneous curvature initial condition (\ref{kconst})
is not appropriate because the far region is asymptotically flat. We
thereby suppose  $A(t=0,r)=B(t=0,r)=1$ and solve the constraint equations
(\ref{hcmp}) and (\ref{mcmp}) to determine  $K(t=0,r)$ and
$K^2_2(t=0,r)$. This treatment is not usually adopted, but it is
suitable for this system because we obtain
\begin{equation}
-{K\over3}\approx-K^2_2\approx\sqrt{{8\pi\over3m_{Pl}^2}
\biggl({\Phi'^2\over2}+{\Phi^2\over r^2}+V\biggr)},
\end{equation}
which approaches zero as $r$ increases. The numerical boundary is
fixed at $r=10\delta$ ($\delta\equiv c\delta_0$).

In regard to the time-evolution of $K^2_2$, we do not have to solve
the equation at the origin which corresponds to (\ref{dotk22dw}),
because in this case we have the relation $K^2_2(t,r=0)=K(t,r=0)/3$
from the regularity condition.

\vskip.5cm
In both the cases (1) and (2), we use a finite difference method with
1000 or 2000 meshes. The Hamiltonian constraint equation (\ref{hcdw})
or (\ref{hcmp}) remains unsolved during the evolution and is used for
checking the numerical accuracy. Through all the calculations the
errors are always less than a few percent.

\vskip.5cm \noindent
(3) Apparent horizon

In our analysis for a global monopole system, we investigate the
global structure of the spacetime, such as the existence of an event
horizon of a black-hole, or the existence of a cosmological horizon.
In numerical relativity, one practical method to see the spacetime
structure is to search an apparent horizon. The apparent horizon is
defined as the outermost/innermost closed 2-surface where the expansion
of a null geodesic congruence vanishes. For the metric (\ref{sphere}),
the expansion, $\Theta_{\pm}$, is written as
\begin{equation}
\Theta_{\pm} =
{k^{2}_{\pm;2}}+{k^{3}_{\pm;3}}=2\bigg[-K^{2}_{2}\pm{(Br)'\over ABr}
\biggl],
\end{equation}
where $k^{\mu}_{\pm}=(-1,\pm A^{-1},0,0)$ is an outgoing $(+)$ or
ingoing $(-)$ null vector.

We observe the signs of $\Theta_{\pm}$ at all points in the numerical
spacetime, and see if there is a 2-surface where $\Theta_+$ or
$\Theta_-$ vanishes. In a black hole system, it is proved that, if an
apparent horizon exists, an event horizon also exists outside (or
coincides with) it. Here, as well as we search for a black-hole
horizon, we use an apparent horizon as a tool for finding a cosmological
horizon; a cosmological horizon exists inside (or coincides with) the
apparent horizon.

%%%%%%%%%%%%%%%%%%%%%%%%%%%%%%%%%%%%%%%%%%%%%%%%%%%%%%%%%%%

\newpage\noindent
\baselineskip = 22pt
{\bf Figure Captions}
\vspace{.3 cm}

\noindent
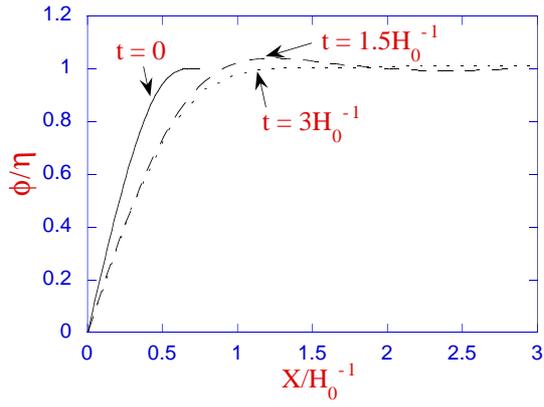
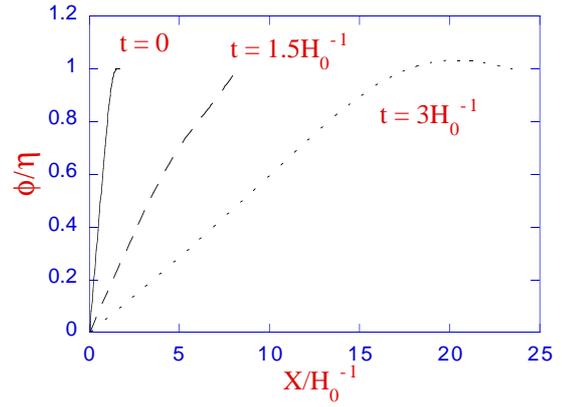
{\bf Fig.1}: Examples of the evolution of a domain wall. We set
$\eta=0.2m_{Pl}$ in (a), $\eta=0.6m_{Pl}$ in (b), and $c=1$ and
$\lambda=0.1$ in both cases. The case (a) expresses a stable wall and
the case (b) expresses an inflating wall. The abscissa is a proper
distance along the $x$-axis from the origin.

\vskip .5cm\noindent
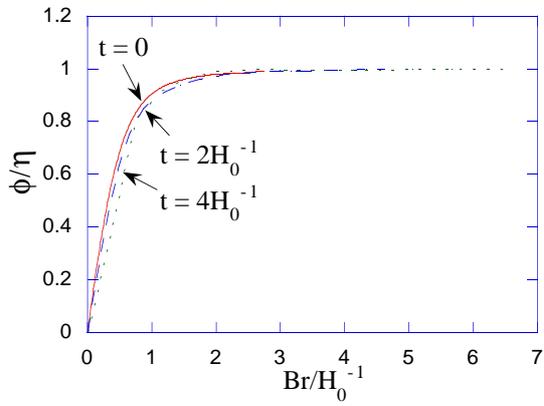
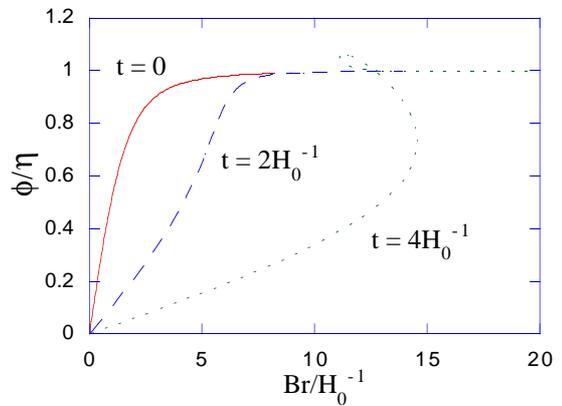
{\bf Fig.2}: Examples of the evolution of a global monopole.
Parameters we choose are the same as in Fig.1. In the case of (b) we
find that the bottle-neck structure appears.

\vskip .5cm\noindent
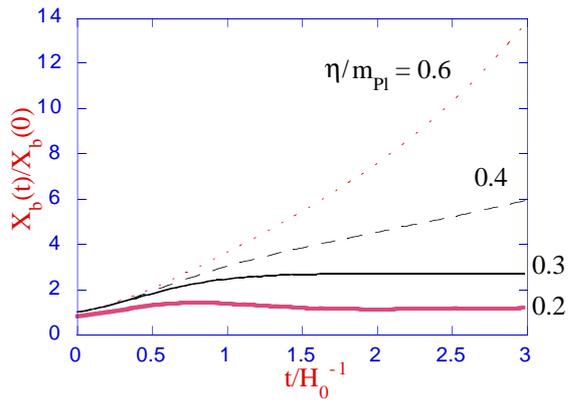
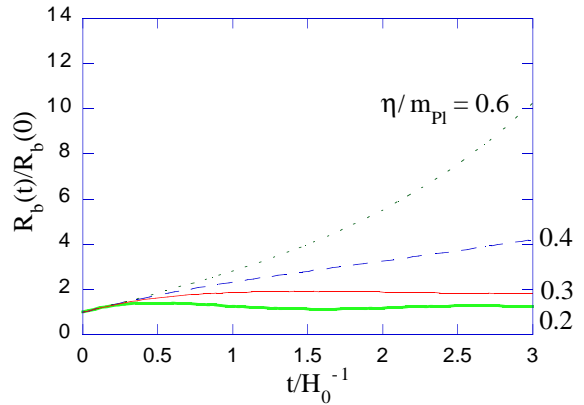
{\bf Fig.3}: Dependence of the evolution of a domain wall and a
monopole on $\eta$. We plot trajectories of the positions of
$\Phi=\eta/2$. We set $\lambda=0.1$ and $c=1$ in both cases. (a) and
(b) express the dynamics of a domain wall and a global monopole,
respectively. A domain wall or a monopole expands if
$\eta\stackrel{>}{\sim}0.33m_{Pl}$.

\vskip .5cm\noindent
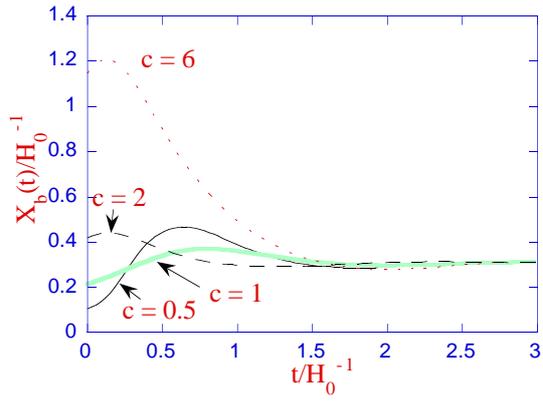
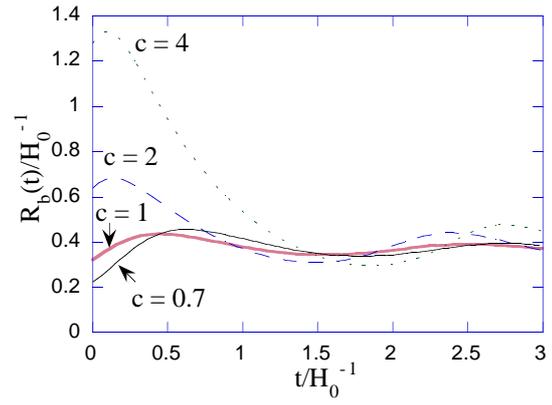
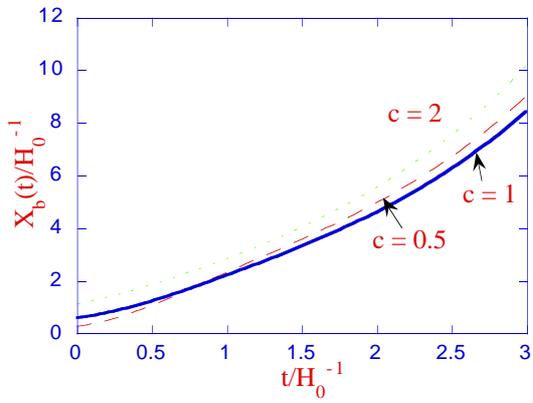
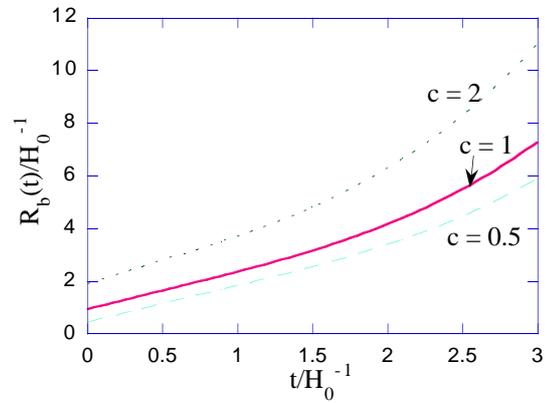
{\bf Fig.4}: Dependence of the evolution on the initial size. We set
$\lambda=0.1$ in all cases. (a): domain wall, $\eta=0.2m_{Pl}$. (b):
global monopole, $\eta=0.2m_{Pl}$. (c): domain wall, $\eta=0.6m_{Pl}$.
(d): global monopole, $\eta=0.6m_{Pl}$. Only $\eta$ determines whether
or not inflation occurs, regardless of the initial size.

\vskip .5cm\noindent
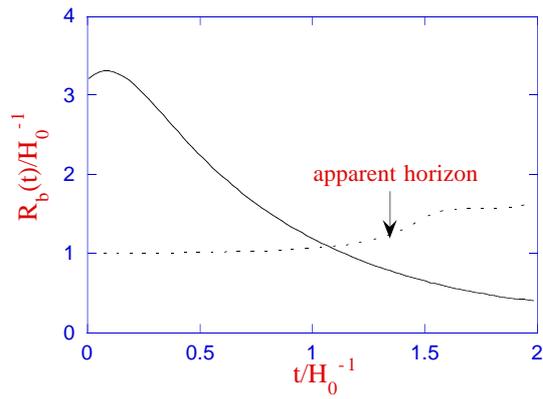
{\bf Fig.5}: Trajectories of the apparent (cosmological) horizon and
of the boundary of a shrinking monopole. We set $c=10$, $\eta=0.2m_{Pl}$
and $\lambda=0.1$. The boundary of the shrinking monopole crosses the
horizon.

\vskip .5cm\noindent
{\bf Fig.6}: Acceleration of test particles in spacetimes (a) with a
global monopole and (b) with a gauge monopole. We set $\lambda=0.1$ in
(a), $\lambda/e^2=0.1$ in (b), and $\eta=0.2m_{Pl}$ in both cases. We
find that the spacetime with a global monopole has a repulsive nature,
while the spacetime with a gauge monopole has an attractive nature.

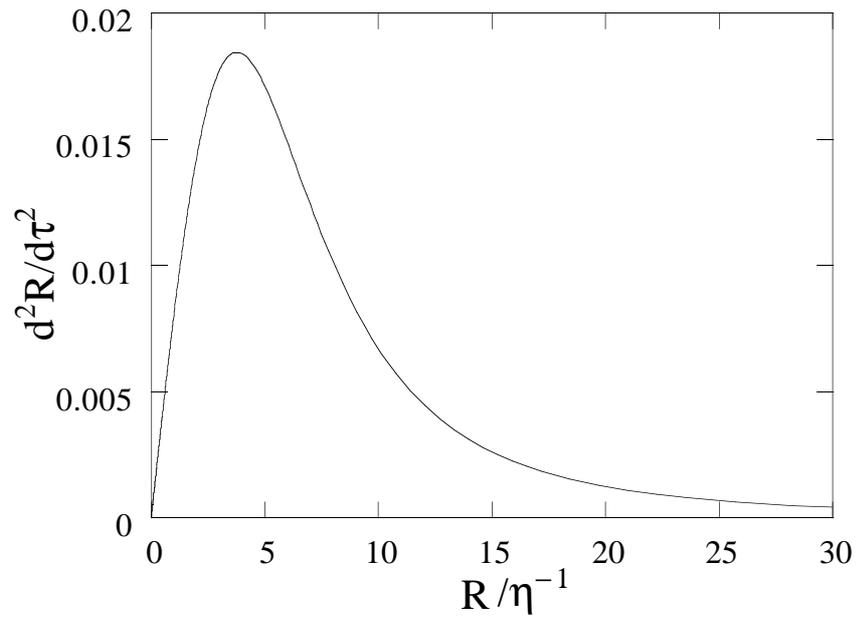
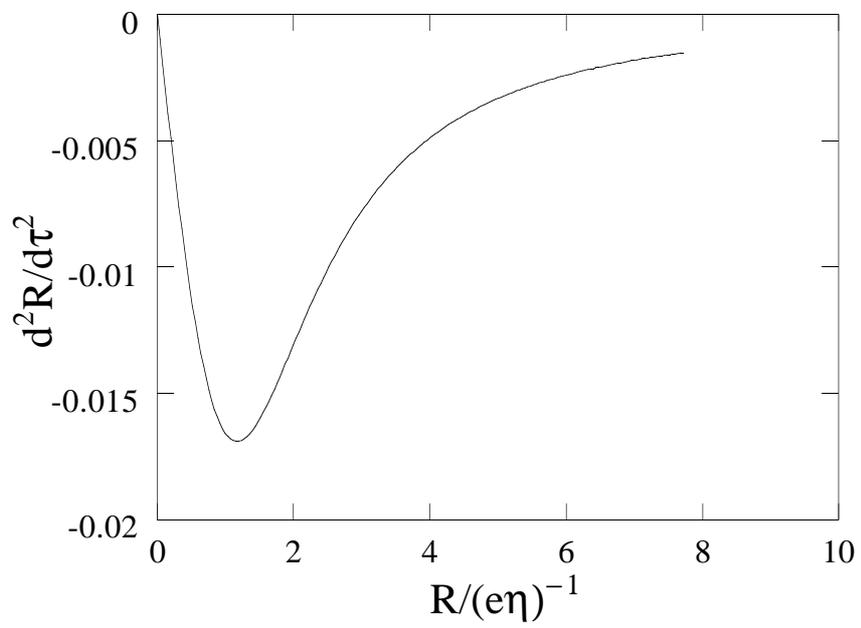
\end{document}